\documentclass[12pt]{article}
 \usepackage{epsfig}
 \def\be{\begin{equation}}
 \def\ee{\end{equation}}
 \def\bea{\begin{eqnarray}}
 \def\eea{\end{eqnarray}}
 \usepackage{graphicx}

 \catcode`\@=11
 \def\lsim{\mathrel{\mathpalette\@versim<}}
 \def\gsim{\mathrel{\mathpalette\@versim>}}
 \def\@versim#1#2{\vcenter{\offinterlineskip
 \ialign{$\m@th#1\hfil##\hfil$\crcr#2\crcr\sim\crcr } }}
 \catcode`\@=12

 \catcode`@=12
\begin{document}
 \thispagestyle{empty}
 \begin{flushright}
 UCRHEP-T602\\
 Oct 2020\
 \end{flushright}
 \vspace{0.6in}
 \begin{center}
 {\LARGE \bf Split Left-Right Symmetry and\\ 
 Scotogenic Quark and Lepton Masses\\}
 \vspace{1.2in}
 {\bf Ernest Ma\\}
 \vspace{0.2in}
{\sl Physics and Astronomy Department,\\ 
University of California, Riverside, California 92521, USA\\}
\end{center}
 \vspace{1.2in}

\begin{abstract}\
A model of split left-right symmetry is proposed, where the first family 
of quarks and leptons transforms under $SU(2)_R$ but the heavier two families 
do not.  The Higgs scalar sector consists only of an $SU(2)_L$ doublet and 
an $SU(2)_R$ doublet.  The $u,d$ quarks and the electron, as well as the 
neutrinos, are massless at tree level, but become massive radiatively 
through their interactions with a simple dark sector.  Verifiable 
consequences include $Z'$ and Higgs properties, neutrino mixing 
and rare decays, etc. 
\end{abstract}

 \newpage
 \baselineskip 24pt
\noindent \underline{\it Introduction}~:~
In the standard model (SM) of particle interactions, neutrinos are massless. 
It is thus reasonable to think that they may become massive through a simple 
mechanism beyond the SM.  In 2006, this was accomplished~\cite{m06} by 
postulating the interactions of neutrinos with dark matter, such that 
Majorana neutrino masses are generated in one loop, i.e. the scotogenic 
mechanism.  Consider now the other fermion masses.  They appear at tree 
level through their interactions with the one Higgs boson.  Yet those of 
the first family, i.e. $m_u,m_d,m_e$, are much smaller than the rest. 
To understand this, a possible approach is the extension of the SM to the 
left-right model, where $(u,d)_R$ and $(\nu_e,e)_R$ are now doublets 
under $SU(2)_R$.  To pair these up with the existing $(u,d)_L$ and 
$(\nu_e,e)_L$, a scalar bidoublet is required.  Suppose this is 
withheld~\cite{dw87,m88,m89,bms03,ms18}, then these fermions are massless 
at tree level. [The analog for neutrino mass is the withholding of a 
scalar triplet~\cite{ms98} in the SM.]  They may then be 
generated in one loop, again using the scotogenic mechanism, as proposed in 
this paper.  As for the heavier two families, they may still acquire masses 
as in the SM, in contrast to a previous proposal that all obtain scotogenic 
masses~\cite{m14}.  This notion that families are split in their left-right 
assignments leads to observable consequences involving $Z'$ and Higgs 
properties, neutrino mixing, rare decays, etc.

\noindent \underline{\it Model of split left-right symmetry}~:~
The particle content of this model is shown in Table~1.  Note that the 
notion of an extra gauge $SU(2)$ under which the quarks and 
leptons do not transform is not new~\cite{bmw81,bmw82}.  Here the first 
family transforms but not the other two. The $Z_2^D$ 
symmetry is imposed to enable scotogenic masses for $u,d,e$ and 4 neutrinos.
It is equivalent~\cite{m15} to assigning $L=-1$ to $\eta_{L,R}$, $B=1/3$ to 
$\zeta_{L,R}$, and $L=B=0$ to $N$, then using $R$ parity, i.e. 
$(-1)^{3B+L+2j}$ for $Z_2^D$.
\begin{table}[tbh]
\centering
\begin{tabular}{|c|c|c|c|c|c|}
\hline
fermion/scalar & $SU(3)_C$ & $SU(2)_L$ & $SU(2)_R$ & $U(1)_X$ & $Z_2^D$ \\
\hline
$(u,d)_L,(c,s)_L,(t,b)_L$ & 3 & 2 & 1 & 1/6 & + \\ 
$(u,d)_R$ & 3 & 1 & 2 & 1/6 & + \\
$c_R,t_R$ & 3 & 1 & 1 & 2/3 & + \\ 
$s_R,b_R$ & 3 & 1 & 1 & $-1/3$ & + \\ 
\hline
$(\nu_e,e)_L,(\nu_\mu,\mu)_L,(\nu_\tau,\tau)_L$ & 1 & 2 & 1 & $-1/2$ & $+$ \\ 
$(\nu_e,e)_R$ & 1 & 1 & 2 & $-1/2$ & + \\
$\mu_R,\tau_R$ & 1 & 1 & 1 & $-1$ & + \\ 
\hline
$\Phi_L=(\phi_L^+,\phi_L^0)$ & 1 & 2 & 1 & 1/2 & + \\ 
$\Phi_R=(\phi_R^+,\phi_R^0)$ & 1 & 1 & 2 & 1/2 & + \\ 
\hline
$\eta_L=(\eta_L^+,\eta_L^0)$ & 1 & 2 & 1 & 1/2 & $-$ \\ 
$\eta_R=(\eta_R^+,\eta_R^0)$ & 1 & 1 & 2 & 1/2 & $-$ \\ 
$\zeta_L = (\zeta_L^{2/3},\zeta_L^{-1/3})$ & 3 & 2 & 1 & 1/6 & $-$ \\ 
$\zeta_R = (\zeta_R^{2/3},\zeta_R^{-1/3})$ & 3 & 1 & 2 & 1/6 & $-$ \\ 
\hline
$N_{1R},N_{2R},N_{3R}$ & 1 & 1 & 1 & 0 & $-$ \\
\hline
\end{tabular}
\caption{Fermion and scalar content of split left-right model.}
\end{table}
The $SU(2)_L \times SU(2)_R \times U(1)_X$ symmetry is broken by 
$\langle \phi_L^0 \rangle = v_L$ and $\langle \phi_R^0 \rangle = v_R$.    
The mass-squared matrix spanning $(W_L^0,W_R^0,B)$ is given by
\begin{equation}
{\cal M}^2 = {1 \over 2} \pmatrix{g_L^2 v_L^2 & 0 & -g_L g_X v_L^2 \cr 
0 & g_R^2 v_R^2 & -g_R g_X v_R^2 \cr -g_L g_X v_L^2 & -g_R g_X v_R^2 & 
g_X^2 (v_R^2 + v_L^2)}.
\end{equation}
Using $e^{-2} = g_L^{-2} + g_R^{-2} + g_X^{-2}$, and 
assuming $g_L=g_R$ with $x = \sin^2 \theta_W$, 
the physical gauge bosons are 
\begin{equation}
\pmatrix{A \cr Z \cr Z'} = \pmatrix{\sqrt{x} & \sqrt{x} & \sqrt{1-2x} \cr 
\sqrt{1-x} & -x/\sqrt{1-x} & -\sqrt{x(1-2x)/(1-x)} \cr 0 & 
\sqrt{(1-2x)/(1-x)} & -\sqrt{x/(1-x)}} \pmatrix{W_L^0 \cr W_R^0 \cr B}.
\end{equation}
The photon $A$ is massless, whereas the $2 \times 2$ mass-squared matrix 
spanning $(Z,Z')$ is given by
\begin{equation}
{\cal M}^2_{Z,Z'} = {e^2 \over 2}\pmatrix{ v_L^2/x(1-x) & 
v_L^2/(1-x)\sqrt{1-2x} \cr v_L^2/(1-x)\sqrt{1-2x} & (1-x) v_R^2/x(1-2x) + 
x v_L^2/(1-x)(1-2x)}.
\end{equation}
The $Z-Z'$ mixing is then about $x \sqrt{1-2x} v_L^2/(1-x)^2 v_R^2$ which is 
constrained experimentally to be less than $10^{-4}$, implying thus 
$v_R > 9.3$ TeV. Whereas $Z$ couples to the current 
\begin{equation}
J_Z = J_{3L}-xJ_{em}
\end{equation}
with strength $e/\sqrt{x(1-x)}$ as in the SM, $Z'$ couples to the current 
\begin{equation}
J_{Z'} = x J_{3L} + (1-x) J_{3R} - x J_{em}
\end{equation}
with strength $e/\sqrt{x(1-x)(1-2x)}$.  Thus neutral flavor changing $Z'$ 
couplings occur through $J_{3R}$.

The charged $W_L^\pm$ and $W_R^\pm$ bosons have masses given by
\begin{equation}
M^2_{W_L} = {e^2 v_L^2 \over 2x}, ~~~ M^2_{W_R} = {e^2 v_R^2 \over 2x}.
\end{equation}
Whereas $W_L^\pm$ couples the $(u,c,t)$ quarks to the $(d,s,b)$ quarks 
through the $3 \times 3$ unitary CKM matrix, $W_R^\pm$ couples only 
one linear combination of $up$ quarks to one linear combination of 
$down$ quarks.

\noindent \underline{\it Scotogenic masses for $u,d,e$ and neutrinos}~:~
The gauge structure of the heavier two families of quarks and leptons is 
identical to that of the SM.  Hence they acquire tree-level masses from 
$\langle \phi_L^0 \rangle = v_L = 174$ GeV as usual.  However, $(u,d)_L$ 
and $(\nu_e,e)_L$ cannot pair up with $(u,d)_R$ and $(\nu_e,e)_R$ at tree 
level because the usual left-right scalar bidoublet is absent.  Nevertheless, 
with the help of the postulated dark sector, they acquire one-loop radiative 
masses as shown in Figures 1 to 4.

\begin{figure}[htb]
\vspace*{-5cm}
\hspace*{-3cm}
\includegraphics[scale=1.0]{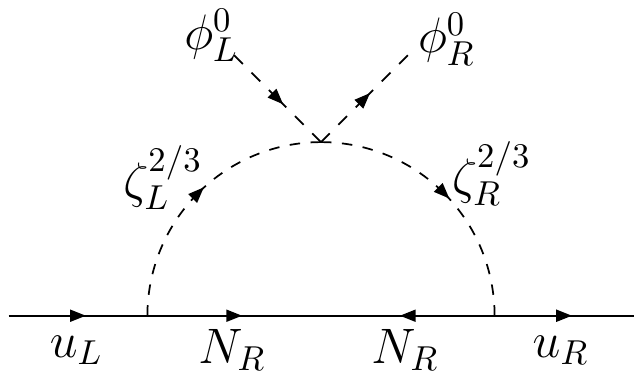}
\vspace*{-21.5cm}
\caption{Scotogenic $u$ quark mass.}
\end{figure}

\begin{figure}[htb]
\vspace*{-5cm}
\hspace*{-3cm}
\includegraphics[scale=1.0]{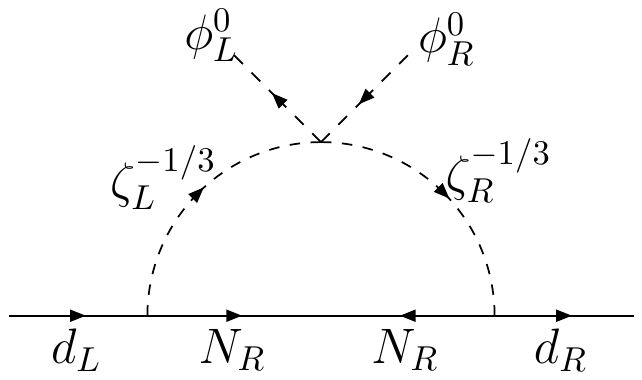}
\vspace*{-21.5cm}
\caption{Scotogenic $d$ quark mass.}
\end{figure}

The most general $3 \times 3$ mass matrix linking $(u,c,t)_L$ to $(u,c,t)_R$ 
may be chosen as
\begin{equation}
{\cal M}_u = \pmatrix{m_{uu} & 0 & 0 \cr m_{cu} & m_{cc} & 0 \cr m_{tu} & 0 & 
m_{tt}},
\end{equation}
where $m_{uu},m_{cu},m_{tu}$ are radiative contributions from Fig.~1.  The 
corresponding $3 \times 3$ mass matrix linking $(d,s,b)_L$ to $(d,s,b)_R$ 
is then of the form
\begin{equation}
{\cal M}_d = \pmatrix{m_{dd} & m_{ds} & 0 \cr m_{sd} & m_{ss} & m_{sb} \cr 
m_{bd} & m_{bs} & m_{bb}},
\end{equation}
where $m_{dd},m_{sd},m_{bd}$ are radiative contributions from Fig.~2. After 
diagonalization by unitary transformations $U_L^\dagger$ on the left and 
$U_R$ on the right, the mismatch $U_L^\dagger (u) U_R(d)$ becomes the well-known 
observed charged-current mixing matrix $U_{CKM}$ as usual.

\begin{figure}[htb]
\vspace*{-5cm}
\hspace*{-3cm}
\includegraphics[scale=1.0]{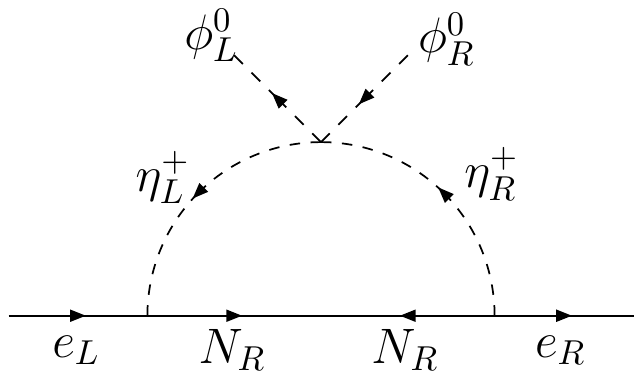}
\vspace*{-21.5cm}
\caption{Scotogenic electron mass.}
\end{figure}
The charged-lepton mass matrix may be organized as~\cite{m98}
\begin{equation}
{\cal M}_e = \pmatrix{m_{ee} & 0 & 0 \cr m_{\mu e} & m_{\mu\mu} & 0 \cr 
m_{\tau e} & 0 & m_{\tau\tau}},
\end{equation}
where $m_{ee},m_{\mu e},m_{\tau e}$ are radiative contributions from Fig.~3.  
As for neutrinos, there are three $\nu_L$s which belong to $SU(2)_L$ doublets 
and one $\nu_R$ which belongs to an $SU(2)_R$ doublet.  They acquire 
scotogenic masses in a $4 \times 4$ matrix.  In Fig.~4, one linear combination 
of the three $\nu_L$s pairs up with the one $\nu_R$ to form a Dirac mass. 

\begin{figure}[htb]
\vspace*{-5cm}
\hspace*{-3cm}
\includegraphics[scale=1.0]{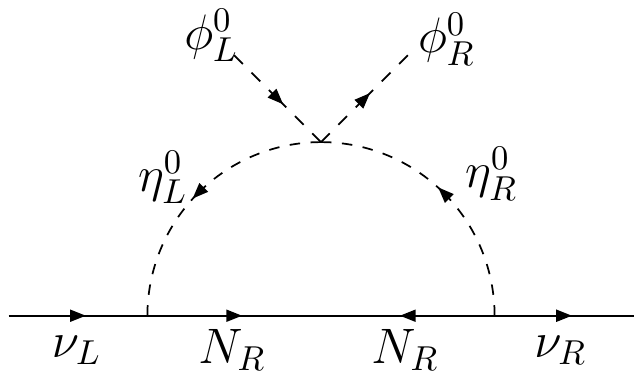}
\vspace*{-21.5cm}
\caption{Scotogenic Dirac neutrino mass.}
\end{figure}
\begin{figure}[htb]
\vspace*{-5cm}
\hspace*{-3cm}
\includegraphics[scale=1.0]{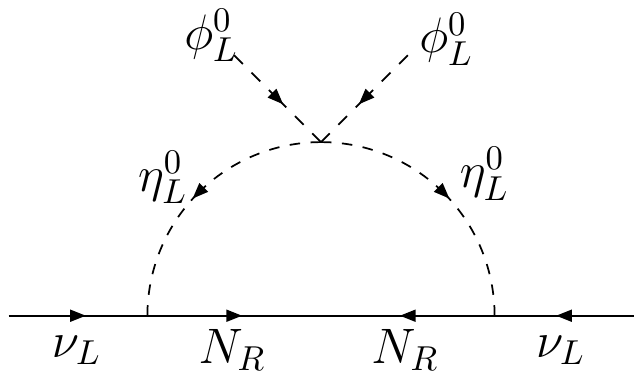}
\vspace*{-21.5cm}
\caption{Scotogenic Majorana $\nu_L$ mass.}
\end{figure}
In Fig.~5, the three $\nu_L$s obtain radiative Majorana masses as in the 
original scotogenic model~\cite{m06}.
\begin{figure}[htb]
\vspace*{-5cm}
\hspace*{-3cm}
\includegraphics[scale=1.0]{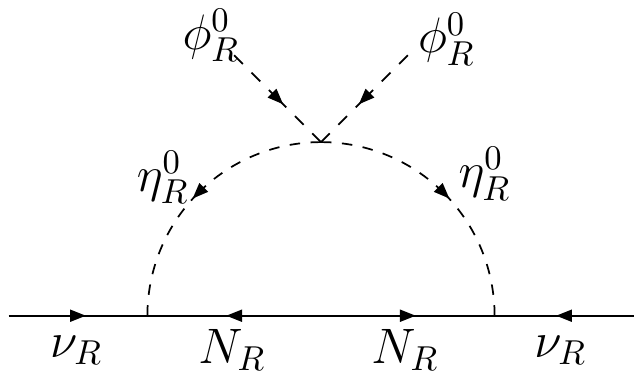}
\vspace*{-21.5cm}
\caption{Scotogenic Majorana $\nu_R$ mass.}
\end{figure}
In Fig.~6, the analog mechanism is used for the one $SU(2)_R$ neutrino 
$\nu_R$.  This mechanism is akin to that~\cite{m96} for 
a singlet neutrino in the SM with an extra gauge $U(1)$.

\noindent \underline{\it New gauge bosons}~:~
All quarks and leptons transform under $SU(2)_L$ as in the SM.  This means 
that the known gauge bosons, $A, W_L^\pm, Z$, have interactions as in the 
SM.  As for the new gauge bosons, $W_R^\pm, Z'$, their interactions are 
different from all previously proposed left-right models, because 
only one copy each of $(u,d)_R$ and $(\nu_e,e)_R$ transforms under $SU(2)_R$. 
Note that these fermions are not mass eigenstates themselves, as shown in 
Eqs.~(7,8,9).  They are linear combinations of mass eigenstates, i.e. 
\begin{equation}
\pmatrix{u \cr d}_R \to \pmatrix{u' \cr d'}_R = \pmatrix{U_R^{uu}(u)u 
+ U_R^{uc}(u)c + U_R^{ut}(u)t \cr  U_R^{dd}(d)d + U_R^{ds}(d)s 
+ U_R^{db}(d)b}_R,
\end{equation}
and similarly for $(\nu_e,e)_R$, except that $(\nu_e)_R$ now has 4 components. 
With the natural expectation that radiative quark and lepton masses are much 
smaller than their tree-level counterparts and using the observed hierarchy 
of masses among families, it is reasonable to find 
$|U_R^{ut}(u)| << |U_R^{uc}(u)| << |U_R^{uu}(u)| \sim 1$ and 
$|U_R^{db}(d)| << |U_R^{ds}(d)| << |U_R^{dd}(d)| \sim 1$, etc. 

The new charged $W_R^\pm$ gauge boson couples $u'_R$ to $d'_R$ and 
$\nu'_R$ to $e'_R$, whereas $Z'$ couples to 
$\overline{u'_R} \gamma u'_R - \overline{d'_R} \gamma d'_R$ and 
$\overline{\nu'_R} \gamma \nu'_R - \overline{e'_R} \gamma e'_R$ 
through $J_{3R}$ in Eq.~(5).  This means that flavor-changing neutral 
currents, absent at tree level in the SM, are now present through $Z'$ 
exchange.  Their effects are however suppressed because of the expected 
small deviations of $u'_R$ from $u_R$, $d'_R$ from $d_R$, $e'_R$ from 
$e_R$, and $\nu'_R$ from $(\nu_e)_R$.

A prime example is the possible rare decay $\mu \to eee$.  Experimentally, 
its branching fraction is less than $1.0 \times 10^{-12}$~\cite{pdg18}.  
Here it is mediated by $Z'$ exchange:
\begin{eqnarray}
{\cal A}(\mu \to eee) &=& {e^2 U_R^{e \mu}(e) \over x(1-x)(1-2x) M^2_{Z'}} 
\bar{e} \gamma^\alpha \left[ -{1\over 2} (1-x) \right] \left( {1+\gamma_5 
\over 2} \right) \mu \nonumber \\ &\times& \bar{e} \gamma_\alpha \left[ 
{x\over 2} \left( {1-\gamma_5 \over 2} \right) + \left( -{1 \over 2} + 
{3x \over 2} \right) \left( {1+\gamma_5 \over 2} \right) \right] e.
\end{eqnarray} 
The resulting constraint~\cite{n93} is
\begin{equation}
[U_R^{e \mu}(e)]^2 \left( {M_Z^2 \over M_{Z'}^2} \right)^2 {(1-x)^2 \over 
4(1-2x)^2} [2-12x+19x^2] < 1.0 \times 10^{-12}.
\end{equation}
For $x = \sin^2 \theta_W = 0.23$, 
\begin{equation}
U_R^{e \mu}(e) \left( { M_Z^2 \over M_{Z'}^2 } \right) < 2.83 \times 10^{-6}
\end{equation}
is obtained.  Using $\mu-e$ conversion in nuclei, this bound may be 
improved~\cite{bnt93} by about a factor of two.  Other rare decays 
sensitive to $Z'$ are $s \to d l^+ l^-$, $b \to d l^+ l^-$, etc.
The present collider bound on $M_{Z'}$ is a few TeV~\cite{pdg18}.

\noindent \underline{\it Exotic Higgs decays}~:~
The Higgs sector of this model is very simple.  The part containing 
$\Phi_{L,R}$ only is given by
\begin{equation}
V_{\Phi} = \mu_L^2 \Phi_L^\dagger \Phi_L + \mu_R^2 \Phi_R^\dagger \Phi_R + 
{1 \over 2} \lambda_L  (\Phi_L^\dagger \Phi_L)^2 + 
{1 \over 2} \lambda_R  (\Phi_R^\dagger \Phi_R)^2 + 
\lambda_{LR} (\Phi_L^\dagger \Phi_L)(\Phi_R^\dagger \Phi_R).
\end{equation}
After spontaneous symmetry breaking with $\langle \phi_L^0 \rangle = v_L$ 
and $\langle \phi_R^0 \rangle = v_R$, the only physical scalar bosons are 
$H_L = \sqrt{2} Re (\phi^0_L)$ and $H_R = \sqrt{2} Re (\phi^0_R)$, with 
$2 \times 2$ mass-squared matrix given by
\begin{equation}
{\cal M}^2_H = \pmatrix{2 \lambda_L v_L^2 & 2 \lambda_{LR} v_L v_R \cr 
2 \lambda_{LR} v_L v_R & 2 \lambda_R v_R^2}.
\end{equation}
In the limit $\lambda_{LR} \to 0$, $H_L$ decouples from $H_R$ and appears 
to be the known SM Higgs boson.  However, because some fermion masses are 
radiative in origin, Higgs decays will not be exactly those of the SM, 
as pointed out in Refs.\cite{fm14,fmz16}.

Consider for simplicity the $2 \times 2$ mass matrix linking 
$(\bar{e}_L,\bar{\mu}_L)$ to $(e_R,\mu_R)$:
\begin{equation}
{\cal M}_{e \mu} = \pmatrix{m_{ee} & 0 \cr m_{\mu e} & m_{\mu \mu}},
\end{equation}
where $m_{ee},m_{\mu e}$ come from Fig.~3.  It is diagonalized on the right 
and left by
\begin{equation}
\pmatrix{\cos \theta_R & \sin \theta_R \cr -\sin \theta_R & \cos \theta_R}, ~~~
\pmatrix{\cos \theta_L & -\sin \theta_L \cr \sin \theta_L & \cos \theta_L},
\end{equation}
where $\tan \theta_R = m_{\mu e}/m_{\mu \mu}$ and 
$\tan \theta_L = (m_e/m_\mu) \tan \theta_R$ with the mass eigenvalues 
$m_e = \cos \theta_R m_{ee}$ and $m_\mu = m_{\mu \mu}/\cos \theta_R$.

In the SM, the Higgs coupling matrix is simply $(\sqrt{2}v_L)^{-1}$ 
times ${\cal M}_{e \mu}$.  After diagonalization, it is of the well-known 
form $(\sqrt{2}v_L)^{-1} [m_e \bar{e}e + m_\mu \bar{\mu}\mu]$.  Here, 
because of the radiative mass, the Higgs coupling matrix is 
modified~\cite{fm14}, and becomes
\begin{equation}
\pmatrix{rm_{ee} & 0 \cr rm_{\mu e} & m_{\mu \mu}} \to 
\pmatrix{rm_e & (r-1)\tan \theta_R m_e \cr (r-1)\sin \theta_R 
\cos \theta_R m_\mu & [1+(r-1)\sin^2 \theta_R]m_\mu}.
\end{equation}
It is clear that for $r=1$, it reduces to the SM result.  Here $r \neq 1$ 
in general and the Higgs decays to $\bar{e}e$ and $\bar{\mu}\mu$ are 
changed.  Furthermore, exotic decays to $\bar{\mu}e$ and $\bar{e}\mu$ 
are possible, contrary to the predictions of the SM. 

Recently, ATLAS reports~\cite{atlas20} an observation of the $\mu^+\mu^-$ 
mode at the level $1.2 \pm 0.6$ relative to the SM prediction.  Also, 
CMS has the result~\cite{cms20} $1.2 \pm 0.4 \pm 0.2$.  This 
leaves much room for $r \neq 1$, but more important would be the observation 
of $\mu^\pm e^\mp$.  Oviously, $\tau$ may replace $\mu$ in the above 
analysis, and modifies the Higgs decay to $\tau^+ \tau^-$ and 
allows the $\tau^\pm e^\mp$ mode.

\noindent \underline{\it Neutrino sector}~:~
There are four neutrinos.  They acquire scotogenic masses as shown in 
Figs.~4 to 6.  The $4 \times 4$ neutrino mass matrix spanning 
$(\bar{\nu}_R,\nu_e,\nu_\mu,\nu_\tau)$ is of the form
\begin{equation}
{\cal M}_\nu = \pmatrix{m_R & m_D & 0 & 0 \cr m_D & m_{ee} & m_{e \mu} & 
m_{e \tau} \cr 0 & m_{e \mu} & m_{\mu \mu} & m_{\mu \tau} \cr 0 & m_{e \tau} 
& m_{\mu \tau} & m_{\tau \tau}}.
\end{equation}
If $m_D$ is very small, then this reduces to the usual case of three active 
Majorana neutrinos, with the possible exception of mixing with a mostly 
sterile singlet neutrino $\nu_S = \bar{\nu}_R$.  This may have relevance 
in models where a fourth neutrino is added to explain recent oscillation 
data~\cite{minib18}.  On the other hand, if both $m_{ee}$ and $m_R$ are 
very small, then $\nu_e$ pairs up with $\nu_R$ to form a mostly Dirac 
fermion, in which case neutrinoless double beta decay~\cite{appec19} would be 
suppressed, but the kinematic mass measurement~\cite{katrin19} of $\nu_e$ 
would succeed.  In either case, the lightest $N$ is a possible dark-matter 
candidate~\cite{bkrz20}.  However, since $N$ has the additional interactions 
of Figs.~1 and 2, the constraint from dark-matter relic abundance is 
relaxed, which in turn makes lepton flavor-changing radiative 
decays~\cite{vy15} less restrictive.  Hence $m_N$ may be much less than 
in the original scotogenic model~\cite{m06} of neutrino masses.  This 
means that the color triplet scalars $\zeta_{L,R}$ may be light enough 
to be produced at the Large Hadron Collider and decay to quarks and $N$, 
in analogy to that of scalar quarks in supersymmetry.

\noindent \underline{\it Concluding remarks}~:~
In an extended gauge model with $SU(2)_R$, the particle content is chosen 
(Table 1) so that all neutrinos as well as the $u,d$ quarks and the electron 
are massless at tree level, whereas the second and third families of quarks 
and charged leptons acquire mass as in the SM.  With the implementation 
of a dark sector, $u,d,e,\nu_e$ obtain one-loop radiative Dirac masses 
through their mutual interactions (Figs.~1 to 4) in analogy to the original 
scotogenic mechanism for the Majorana masses of the three left-handecd 
neutrinos (Fig.~5) and similarly for the one right-handed neutrino 
(Fig.~6).  This model of radiative masses for the lightest known particles 
has verifiable consequences in the appearance of $W_R^\pm$ and $Z'$ gauge 
bosons with flavor nondiagonal interactions.  Rare decays such as 
$\mu \to eee$ through $Z'$ exchange are predicted, as well as exotic Higgs 
decays such as $\tau^\pm e^\mp$ and deviations from SM predictions in the 
$\tau^+ \tau^-$, $\mu^+ \mu^-$ and $e^+ e^-$ modes.

\noindent \underline{\it Acknowledgement}~:~
This work was supported in part by the U.~S.~Department of Energy Grant 
No. DE-SC0008541.

\baselineskip 20pt

\bibliographystyle{unsrt}

\begin{thebibliography}{99}
\bibitem{m06} E. Ma, Phys. Rev. {\bf D73}, 077301 (2006).
\bibitem{dw87} A. Davidson and K. C. Wali, Phys. Rev. Lett. {\bf 59}, 393 
(1987).
\bibitem{m88} R. N. Mohapatra, Phys. Lett. {\bf B201}, 517 (1988).
\bibitem{m89} E. Ma, Phys. Rev. Lett. {\bf 63}, 1042 (1989).
\bibitem{bms03} B. Brahmachari, E. Ma, and U. Sarkar, Phys. Rev. Lett. 
{\bf 91}, 911801 (2003).
\bibitem{ms18} E. Ma and U. Sarkar, Phys. Lett. {\bf B776}, 54 (2018).
\bibitem{ms98} E. Ma and U. Sarkar, Phys. Rev. Lett. {\bf 80}, 5716 (1998).
\bibitem{m14} E. Ma, Phys. Rev. Lett. {\bf 112}, 091801 (2014).
\bibitem{bmw81} V. D. Barger, E. Ma, and K. Whisnant, Phys. Rev. Lett. 
{\bf 46}, 1501 (1981).
\bibitem{bmw82} V. D. Barger, E. Ma, and K. Whisnant, Phys. Rev. {\bf D25}, 
1384 (1982).
\bibitem{m15} E. Ma, Phys. Rev. Lett. {\bf 115}, 011801 (2015).
\bibitem{m98} E. Ma, Phys. Lett. {\bf B442}, 238 (1998).
\bibitem{m96} E. Ma, Mod. Phys. Lett. {\bf A11}, 1893 (1996).
\bibitem{pdg18} M. Tanabashi {\it et al.} (Particle Data Group), Phys. Rev. 
{\bf D98}, 030001 (2018).
\bibitem{n93} E. Nardi, Phys. Rev. {\bf D48}, 1240 (1993).
\bibitem{bnt93} J. Bernabeu, E. Nardi, and D. Tommasini, Nucl. Phys. 
{\bf B409}, 69 (1993).
\bibitem{fm14} S. Fraser and E. Ma, Europhys. Lett. {\bf 108}, 11002 (2014). 
\bibitem{fmz16} S. Fraser, E. Ma, and M. Zakeri, Phys. Rev. {\bf D93}, 
115019 (2016).
\bibitem{atlas20} G. Aad {\it et al.} (ATLAS Collaboration), arXiv:2007.07830 
[hep-ex].
\bibitem{cms20} A. M. Sirunyan {\it et al.} (CMS Collaboration), 
arXiv:2009.04353 [hep-ex].
\bibitem{minib18} A. A. Aguilar-Arevalo {\it et al.} (MiniBooNE 
Collaboration), Phys. Rev. Lett. {\bf 121}, 221801 (2018).
\bibitem{appec19} A. Giuliani {\it et al.}, arXiv:1910.04688 [hep-ex].
\bibitem{katrin19} M. Aker {\it et al.} (KATRIN Collaboration), Phys. Rev. 
Lett. {\bf 123}, 221802 (2019).
\bibitem{bkrz20} T. de Boer, M. Klasen, C. Rodenbeck, and S. Zeinstra, 
Phys. Rev. {\bf D102}, 051702 (2020).
\bibitem{vy15} A. Vicente and C. E. Yaguna, JHEP {\bf 1502}, 144 (2015).
\end{thebibliography}

\end{document}